\documentclass[preprint,showpacs,showkeys]{revtex4}
\usepackage{epsf, epsfig, latexsym}
\usepackage{graphicx}
\usepackage{amssymb}
\usepackage{amsmath}
\usepackage{bm}
\usepackage{bbm}
\usepackage{dcolumn}
\usepackage{multirow}
\usepackage{pifont}
\usepackage{calc}
\begin{document}
\bibliographystyle{apsrev}
\title[]
{Parity Odd Domain Structure with Generalized $\theta$ Vacuum }
\author{Eun-Joo \surname{Kim}}
\email[]{ejkim@jbnu.ac.kr}
\author{Jong Bum \surname{Choi}}
\email[]{jbchoi@jbnu.ac.kr}
\affiliation{Division of Science Education,
Chonbuk National University, Jeonju 561-756, Korea}
%
%
\date{\today}
%
%
\begin{abstract}
%
Recent experiments in heavy ion collisions have shown the possibility of creating 
parity-odd domains resulting from the $\theta$ term in strong interaction Lagrangian.
The $\theta$ term originates from the nontrivial solution of QCD vacuum known as the $\theta$ vacuum,
and the value of $\theta$ is taken to be a function of spacetime coordinates in the parity-odd domains.
This means that we have to consider different theories at each point 
so that we need to devise a new approach to define the QCD vacuum.
In this Letter, we suggest a method to generalize the $\theta$ vacuum 
by exploiting the dimension 2 condensates and to calculate the parity-odd domain structure
as the union of gauge slices defined by the constant value of dimension 2 condensate.
\end{abstract}
%
%

\maketitle

%
%
In heavy ion collisions it has been reported that metastable domains 
leading to P and CP violations are observed as a realization of 
an excited vacuum domain~\cite{R1}.
These metastable domains are described as ``P-odd bubbles''
where the parameter $\theta$ introduced as a conjugate variable to the
integral of the topological charge density becomes non-zero.
In contrast to the stringent limit $\theta < 3 \times 10^{-10}$
obtained from the measurement of neutron electric dipole moment~\cite{R2},
the parity violating parameter $\theta$ measured in heavy ion collisions
turns out to be of order $10^{-2}$. 
The large difference between these measurements cannot be easily 
accounted for without introducing new idea to the definition of
quantum chromodynamic (QCD) vacuum.

The variation of the value of $\theta$ up to the order of $10^{8}$ can be
assigned to the existence of different $\theta$-worlds~\cite{R3} generated by instantons
which induce tunnelling from one vacuum to a gauge-rotated vacuum.
One possible explanation of the large variation of $\theta$ between 
the hadronic phase and the quark-gluon plasma phase could be the formation of instanton liquid 
in the hot and dense matter created in heavy ion collisions.
Since the average size of an instanton is taken to be about $\frac{1}{3}$ fermi~\cite{R4},
the formation of instanton liquid needs at least 2 or 3 fermi size domain 
which is rare in hadronic phase.
The formation of large size domain is induced by the fusion process of the colliding hadrons
and we need to devise new method to describe these changes of vacuum domain.
In this Letter, we will give a general idea on the construction of topological spaces of
gluonic vacuum domain and introduce appropriate measure 
for the description of domain structure.
The characteristics of gluonic vacuum domain can be represented by the value of $\theta$
or equivalently by the value of dimension 2 condensate $\langle A_{\mu}^{2} \rangle$~\cite{R5}.

The relation between dimension 2 condensate and instanton contribution has been
confirmed by lattice calculations. The existence of dimension 2 condensate can be checked
by considering the two-point correlation function compared with the lattice gluon propagator.
Quantitative estimation of instanton contribution to dimension 2 condensate can be carried out
through the instanton shape recognition procedure~\cite{R6} in which the topological charge density
\begin{equation}
  Q = \frac{g^2}{32\pi^2} \int d^4 x F^{a}_{\mu\nu} \tilde{F}^{\mu\nu}_{a}
\label{eq1}
\end{equation}
is compared with the lattice one
\begin{equation}
  Q_{\mbox{\small{latt}}}(x) = \frac{1}{2^9 \pi^2} \sum \tilde{\epsilon}_{\mu\nu\rho\sigma}
                       \text{Tr}[\Pi_{\mu\nu}(x)\Pi_{\rho\sigma}(x)], 
\label{eq2}
\end{equation}
where $ \tilde{\epsilon}_{\mu\nu\rho\sigma}$ is the antisymmetric tensor and 
$\Pi_{\mu\nu}(x)$ is the field tensor defined on the lattice.
In this way we can measure the radius of the identified instanton and count
the numbers $n_{I}$ of instantons and $n_{A}$ of anti-instantons.
Then we get
\begin{equation}
  \langle A_{\mbox{\small{inst}}}^2 \rangle 
      = \frac{n_I + n_A}{V} \int d^4 x \sum_{\mu, a} A_{\mu}^{(I)a}(x)  A_{\mu}^{(I)a}(x),
\label{eq3}
\end{equation}
assuming that the QCD vacuum is approximated by the ensemble of non-interacting instantons.
The estimated result is consistent with the one obtained by operator product expansion
so that we can conclude that the instanton liquid picture is useful in deducing the
value of dimension 2 condensate for the long range region with nonperturbative interactions.

The ordinary $\theta$ vacuum is defined by the eigenstate of the gauge transformation
performed between different vacuum states fixed by the gauge field components 
with different topological charges. The topological charge density can be added to the
Lagrangian and then the strong interaction Lagrangian becomes
\begin{equation}
    \mathcal{L} = -\frac{1}{4}F^{a}_{\mu\nu} F^{\mu\nu}_{a}
                  - \theta \frac{g^2}{32\pi^2} F^{a}_{\mu\nu} \tilde{F}^{\mu\nu}_{a}
                  + \sum_{f} \bar{\psi}_{f} \Big[ i\gamma^{\mu}(\partial_\mu - igA_{\mu}) - m_f \Big] \psi_f ,
\label{eq4}
\end{equation}
where $\theta$ is the parameter characterizing the vacuum state.
Effectively the $\theta$-term can be transferred into the mass term of up quark by using
axial anomaly and the results are~\cite{R7}
\begin{equation}
    \mathcal{L}_\theta = -m \cos\theta(\bar{u}_L u_R + \bar{u}_R u_L) 
                        - im\sin\theta(\bar{u}_L u_R - \bar{u}_R u_L) 
\label{eq5}
\end{equation}
representing the flip of handedness in the quark field. There existed a stringent limit
$\theta < 3 \times 10^{-10}$ from the measurement of neutron electric dipole moment,
however, the observations of parity-odd domains in relativistic heavy ion collisions 
by STAR Collaboration and by ALICE Collaboration give strong support 
for the metastable state with the value of $\theta$ in the order of $10^{-2}$.
These large differences in the value of $\theta$ imply that the metastable state has to be
localized in space and time~\cite{R8} and the vacuum domain has to be characterized by 
$\theta = \theta(\mathbf{x},t)$.
This spacetime dependence of $\theta$ can be viewed from a different point of view 
when we classify the points of the vacuum domain by the conditions 
$\theta(\mathbf{x},t) = \theta_i$ with fixed value of $\theta_i$.
The classified points form a set of surfaces in the vacuum domain and the
time evolution of the surfaces generates the unstability of the domain.
For a fixed $\theta_i$, the instanton contributions can be estimated by the value of
dimension 2 condensate $\langle A_{\mu}^{2} \rangle$, and for another $\theta'_i$
we can assign another value of $\langle A_{\mu}^{2} \rangle$.
This situation can be represented by
\begin{equation}
   \langle A_{\mu}^{2} \rangle_\theta = C_\theta
\label{eq6}
\end{equation}
with different $C_\theta$ for each $\theta$. 
Then the spacetime dependence of $\theta$ is naturally transferred into that of
$\langle A_{\mu}^{2} \rangle$, that is~\cite{R9}
\begin{equation}
   \langle A_{\mu}^{2} \rangle = C(\mathbf{x},t),
\label{eq7}
\end{equation}
and for a given $C_\theta$ the points satisfying the condition
$C(\mathbf{x},t) = C_\theta$ form a set corresponding to the given
$\theta$ vacuum. The whole metastable domain can be described by the
collections of these sets representing different $\theta$ vacua.

The change of viewpoints from collections of different $\theta$ vacua into the
sets of gauge slices defined by the value of dimension 2 condensate
$\langle A_{\mu}^{2} \rangle$ gives a good chance to construct a model for gluonic domains. 
The gluonic domains appearing in the metastable state formed by heavy ion collisions
are mainly controlled by the positions of quarks and antiquarks which behave as sources
and sinks of the gluons mediating the strong interactions.
Since the gluonic domains can be combined or divided according to the movements 
of quarks and antiquarks, we can introduce the union and the intersection operations
on the open sets assigned to the gluonic domains.
These assignments can be summarized as~\cite{R10} :
\newcommand{\squishlist}{
 \begin{list}{$\bullet$}
  { \setlength{\itemsep}{0pt}
     \setlength{\parsep}{0pt}
     \setlength{\topsep}{2pt}
     \setlength{\partopsep}{0pt}
     \setlength{\leftmargin}{1.5em}
     \setlength{\rightmargin}{1.5em}
     \setlength{\labelwidth}{1em}
     \setlength{\labelsep}{0.5em} } }
\newcommand{\squishend}{
  \end{list}  }
\squishlist
\item Open sets are the gluonic domains.
\item The union of gluonic domains is a gluonic domain.
\item The intersection between a connected gluonic domain and 
      disconnected gluonic domains is the reverse operation of the union.
\squishend
Now we can construct the topological spaces of gluonic domains classified by the numbers
of quarks and antiquarks existing inside the given gluonic domain.
If we represent the gluonic domain with $a$ quarks and $b$ antiquarks as $D_{a,\bar{b}}$,
then the topological space encompassing $i$ baryons and $j$ antibaryons becomes
\begin{equation}
   T_{i,\bar{j}}~ = ~\{ \phi, ~D_{3,\bar{0}}^{i}D_{0,\bar{3}}^{j},
   ~D_{3,\bar{0}}^{i-1}D_{0,\bar{3}}^{j-1}D_{2,\bar{2}},
   ~D_{3,\bar{0}}^{i-2}D_{0,\bar{3}}^{j-2}D_{2,\bar{2}}^{2},
   \cdots \}.
\label{eq8}
\end{equation}
For example, the space with 3 baryons and 1 antibaryon is given by
\begin{equation}
   T_{3,\bar{1}}~ = ~\{ \phi, ~D_{3,\bar{0}}^{3}D_{0,\bar{3}}, 
                        ~D_{3,\bar{0}}^{2}D_{2,\bar{2}},
                        ~D_{3,\bar{0}}D_{4,\bar{1}}, ~D_{6,\bar{0}} \}.
\label{eq9}
\end{equation}
The last domain $D_{6,\bar{0}}$ represents the case of 6 quarks and
we can find that this domain is divided into 3 baryon domains and
1 antibaryon domain through fragmentation processes.
During the fragmentation processes, any number of meson domains
$D_{1,\bar{1}}$ can be created and these domains can be added to the
classified domains shown in Eq.~(\ref{eq9})~\cite{R10}.
The creation process is affected by the non-zero value of $\theta$ assigned
to the original metastable domain formed by the strong collision of heavy nuclei.

In order to calculate the structures of metastable domain related to the 
spacetime dependent $\theta$, we need to introduce a systematic measure defined
between the positions of quarks existing in the domain.
One of the most general measure can be assigned to the amplitude
defined by the nonlocal condensate~\cite{R11}
\begin{equation}
 \langle : \bar{q}(x)U(x,0)q(0) : \rangle \equiv \langle : \bar{q}(0)q(0) : \rangle Q(x^2),
\label{eq10}
\end{equation}
where $U(x,0)$ represents the connection through the gluonic domain.
Since the gluonic domain is characterized by the value of
$\langle A_{\mu}^{2} \rangle$ at each point, we can relate this value
to the function of $Q(x^2)$ by assuming
\begin{eqnarray}
 \langle : \bar{q}(x)U(x,y)A^{a}_{\mu}(y)A^{\mu}_{a}(y)U(y,0)q(0) : \rangle \propto
 \langle : \bar{q}(x)U(x,y)q(y)\bar{q}(y)U(y,0)q(0) : \rangle , 
\label{eq11}
\end{eqnarray}
which implies the proportionality of the value of $\langle A_{\mu}^{2} \rangle$ to the
probability amplitude to have a quark pair at that point.
The functional form of $Q(x^2)$ can be deduced by introducing a measure
$\mathfrak{M}(Q)$ with the condition
\begin{equation}
  \mathfrak{M}(Q)~~\mbox{decreases ~as} ~~Q~~ \mbox{increases}.
\label{eq12}
\end{equation}
The second condition can be stated for two independent $Q_1$ and $Q_2$ as
\begin{equation}
  \mathfrak{M}(Q_{1}) + \mathfrak{M}(Q_{2}) = \mathfrak{M}(Q_{1}Q_{2}).
\label{eq13}
\end{equation}
Then we get the solution 
\begin{equation}
  \mathfrak{M}(Q)  = -k\ln\frac{Q}{Q_{0}},
\label{eq14}
\end{equation}
where $Q_0$ is a normalization constant and $k$ is an appropriate parameter.
If we try to represent the measure $\mathfrak{M}(Q)$ as a metric function of the
distance between the quark pair, it is possible to write the form of $Q$ as~\cite{R10}
\begin{equation}
  Q = \frac{Q_0}{r^\beta} \exp \left \{ -\frac{1}{k} \frac{r^2 - r }{\ln r} \right\},
\label{eq15}
\end{equation}
where $r = \frac{1}{\ell} | \mathbf{x}- \mathbf{y} |$ with $\ell$ being a scale parameter,
and $\beta$ represents the singular behavior near the quark.
%
 \begin{figure}[htb]
 \includegraphics[width=0.76\linewidth]{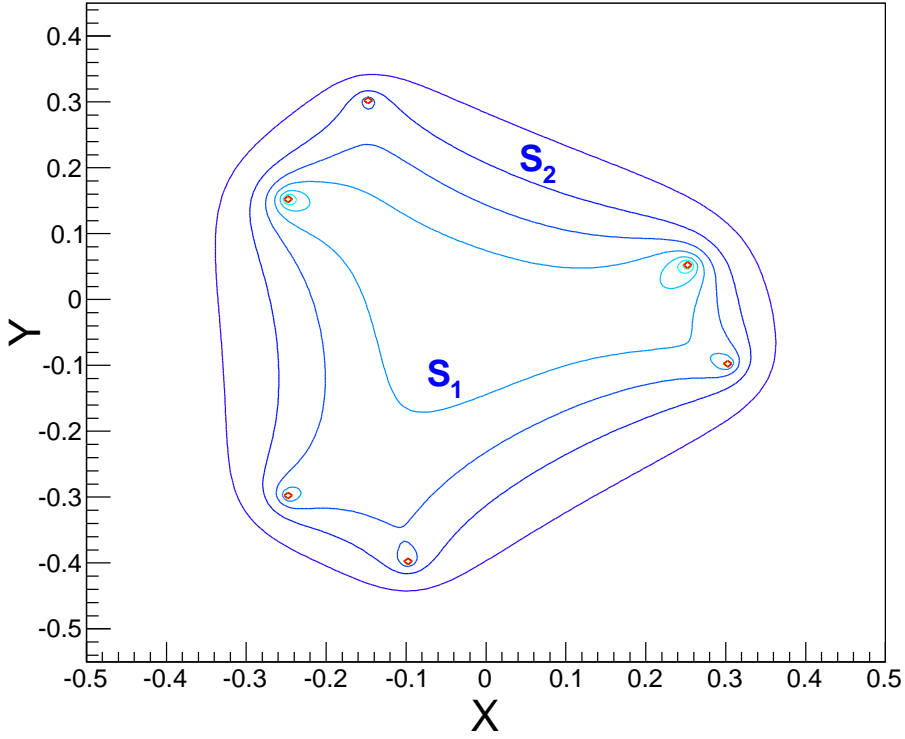}	
\caption{(Color online) 
          Profiles of gauge slices for 6 quark domain represented by $S_1$ and $S_2$.
          The quarks are at $(-0.15, 0.3, 0.0)$, $(-0.25, 0.15, 0.0)$, $(-0.25, -0.3, 0.0)$,
          $(-0.1, -0.4, 0.0)$, $( 0.3, -0.1, 0.0)$, and $(0.25, 0.05, 0.0)$ 
          with $\beta = 1.0$ and $k = 1.0$. The collision axis is the $z$-axis. }
 \label{fig1}
 \end{figure}
In the case of 6 quark domain, the value of $\langle A_{\mu}^{2} \rangle$ 
at the point $\mathbf{x}$ becomes
$
$
\begin{eqnarray}
  \langle A_{\mu}^{2} \rangle 
 &=& A_0^2 \prod_{i=1}^{6} |\mathbf{x}-\mathbf{r}_i|^{-\beta}
     \exp \{ -\frac{1}{k} \frac {|\mathbf{x}-\mathbf{r}_i |^{2} - |\mathbf{x}-\mathbf{r}_i|}
                                  {\ln |\mathbf{x}-\mathbf{r}_i|} \} \\  \nonumber
 && \cdot \Big[ \sum_{i=1}^{6} \prod_{\mathbf{r}_j, \mathbf{r}_k \neq \mathbf{r}_i} |\mathbf{r}_j-\mathbf{r}_k|^{-\beta}
        \exp \{ -\frac{1}{k} \frac { |\mathbf{r}_j-\mathbf{r}_k|^{2} - |\mathbf{r}_j-\mathbf{r}_k|}
                                  {\ln |\mathbf{r}_j-\mathbf{r}_k|}  \}  \\  \nonumber
  && + \sum_{\mathbf{r}_j, \mathbf{r}_k}|\mathbf{r}_j-\mathbf{r}_k|^{-\beta}
        \exp  \{ -\frac{1}{k} \frac { |\mathbf{r}_j-\mathbf{r}_k|^{2} - |\mathbf{r}_j-\mathbf{r}_k|}
                                  {\ln |\mathbf{r}_j-\mathbf{r}_k|} \} \\  \nonumber
  &&\cdot \prod_{\mathbf{r}_\alpha, \mathbf{r}_\gamma \neq \mathbf{r}_j, \mathbf{r}_k} 
                               |\mathbf{r}_\alpha-\mathbf{r}_\gamma|^{-\beta}
       \exp \{ -\frac{1}{k} \frac { |\mathbf{r}_\alpha-\mathbf{r}_\gamma|^{2} - |\mathbf{r}_\alpha-\mathbf{r}_\gamma|}
                                  {\ln |\mathbf{r}_\alpha-\mathbf{r}_\gamma|} \} \Big],
\label{eq16}
\end{eqnarray}
%
where $\mathbf{r}_i$ are the positions of the 6 quarks and $A_0^2$ is a normalization factor.
The first term in the square bracket corresponds to the amplitude to have a meson
and a changed 6 quark structure after quark pair creation at $\mathbf{x}$, 
and the second term represents the amplitude to have a baryon and a pentaquark structure~\cite{R12}.
The calculated results are shown in Fig.~\ref{fig1}.
%
%
 \begin{figure}[htb]
 \includegraphics[width=0.76\linewidth]{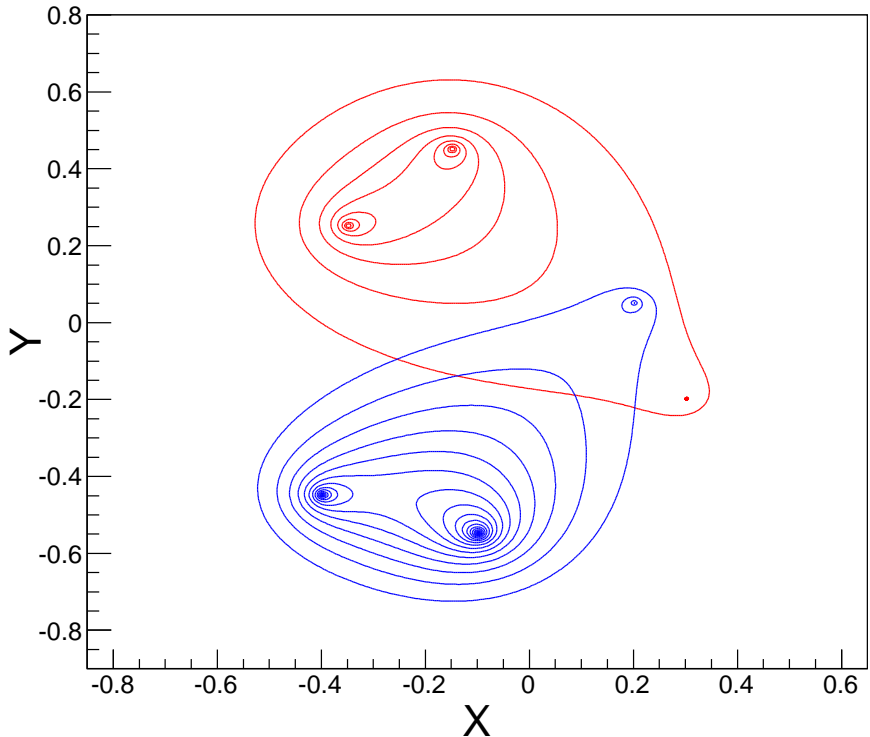}
 \caption{ (Color online) Two baryon domains merging with displaced quarks.}	
 \label{fig2}
 \end{figure}

In  Fig.~\ref{fig2}, we have given the two baryon domains
when they are colliding in the direction of $z$-axis. 
During the collision the domains are expected to change 
according to the movements of quarks.
The final formation of 6 quark domain~\cite{R13} could be processed via one quark pair creation 
forming a baryon and a pentaquark domains and then the union of domains 
with quark pair annihilation leads to the larger structure.

The main difference between the hadronic phase and the quark-gluon plasma phase is 
the range of strong interactions. Since the strong interactions are mediated 
by the propagation of gluons, it is important to check the quantization process 
leading to the definition of gluon propagator. 
In path-integral quantization, the functional integration cannot be carried out
without the gauge fixing procedure.
The well known approach proposed by Faddeev and Popov~\cite{R14} is to factorize the integration space 
into the volume of the orbit traced by gauge group and some surface that
intersects the gauge orbit only once. 
The equation which fixes the integration surface is called
the gauge condition and in perturbation theory this condition is usually given by
constant $\langle A_{\mu}^{2} \rangle$.
However, in nonperturbative region, the value of $\langle A_{\mu}^{2} \rangle$ has to be
position-dependent because the condensate values are taken to be non-zero 
only within the region where the quarks and gluons interact~\cite{R15}.
Then we can define the surfaces $S_1$ and $S_2$ by 
$\langle A_{\mu}^{2}(x) \rangle = C_1$ and
$\langle A_{\mu}^{2}(x) \rangle = C_2$ as in Fig.~\ref{fig1}.
The gauge field $\mathbf{A}_{\mu}$ can be quantized on these gauge slices and 
we can find that the gluons can propagate long distance in the nonperturbative region.
The relation between $S_1$ and $S_2$ can be deduced from the equation~\cite{R16}
\begin{equation}
 \mathbf{A}^{g'}_{i} = \Big ( \delta_{ij}
      - \nabla_i \frac{1}{\nabla^2}\nabla_j \Big ) \mathbf{A}_j + O(A^2_\mu),
\label{eq17}
\end{equation}
where the gauge transformed field $\mathbf{A}^{g'}_{i}$ is represented 
in terms of original $\mathbf{A}_j$.
In traditional approaches we neglect the $O(A^2_\mu)$ terms,
but the $O(A^2_\mu)$ terms cannot be neglected when the dimension 2 condensate
$\langle A_{\mu}^{2} \rangle$ exists and the volume of the gauge orbit becomes
dependent on the value of $A_{\mu}^2$.
This situation can be interpreted such that the surfaces $S_1$ and $S_2$ are
the gauge slices fixed by constant values of $\langle A_{\mu}^{2} \rangle$
and these gauge slices are related by field-dependent gauge transformations
inducing new picture of nonperturbative QCD vacuum.

In summary, we have tried to generalize the $\theta$ vacuum by exploiting 
the dimension 2 condensates and to introduce gluonic vacuum domains 
as the sets of gauge slices defined by the constant value of $\langle A_{\mu}^{2} \rangle$.
We can construct the topological spaces of gluonic domains and the functional form of
$\langle A_{\mu}^{2} \rangle$ has been deduced from the measure assigned to the amplitude
of nonlocal quark condensate. The calculated 6 quark domain is large enough to encompass
instanton liquid so that the observed value of order $10^{-2}$ for the parity violating
parameter $\theta$ can be explained in contrast to the limit
$\theta < 3 \times 10^{-10}$ obtained from neutron data.
The effects of gluon propagation on the gauge slices extended over 
the whole gluonic domain need further study.

%
%
\begin{acknowledgments}
%
%
This work was supported in part by research funds of the LG Yonam Foundation.
\end{acknowledgments}
%
%
{}
%
%
%
\end{document}